\documentclass[conference]{IEEEtran}
\pdfoutput = 1
\IEEEoverridecommandlockouts
\usepackage{cite}
\usepackage{amsmath,amssymb,amsfonts,amsthm}
\usepackage{algorithmic}
\usepackage{graphicx}
\usepackage{listings}
\usepackage{textcomp}
\usepackage{xcolor}
\usepackage{verbatim}
\usepackage{siunitx}
\usepackage[utf8]{inputenc}
\usepackage{lipsum}
\usepackage{booktabs}

\usepackage{xspace}
\usepackage[colorlinks,urlcolor=black,linkcolor=black,citecolor=black,linktocpage=true]{hyperref}
\usepackage{doi}

\newcommand{\figref}[1]{\figurename~\ref{#1}}

\definecolor{dkred}{RGB}{87,10,10}

\lstdefinelanguage[x]{sql}[]{sql}{morekeywords={WITH,DOUBLE,REGEXP_EXTRACT,FROM_UTF8,DAY_OF_WEEK,TO_BASE_64}}

\makeatletter
\newcommand\nocaption{\renewcommand\p@subfigure{}
    \renewcommand\thesubfigure{\thefigure\alph{subfigure}}
}
\makeatother

\title{Dependency Solving Is Still Hard,\\ but We Are Getting Better at It}

\author{\IEEEauthorblockN{Pietro Abate}
  \IEEEauthorblockA{Nomadic Labs \\
    Paris, France \\
    pietro.abate@nomadic-labs.com
  }
  \and
  \IEEEauthorblockN{Roberto Di Cosmo}
  \IEEEauthorblockA{Inria and Université de Paris \\
    Paris, France \\
    roberto@dicosmo.org
  }
  \and
  \IEEEauthorblockN{Georgios Gousios}
  \IEEEauthorblockA{Delft Univ.~of Technology \\
    Delft, The Netherlands \\
    g.gousios@tudelft.nl
  }
  \and
  \IEEEauthorblockN{Stefano Zacchiroli}
  \IEEEauthorblockA{Université de Paris and Inria \\
    Paris, France \\
    zack@irif.fr
  }
}

\begin{document}
\maketitle

\begin{abstract}
  Dependency solving is a hard (NP-complete) problem in all non-trivial
  component models due to either mutually incompatible versions of the same
  packages or explicitly declared package conflicts. As such, software upgrade
  planning needs to rely on highly specialized dependency solvers, lest falling
  into pitfalls such as incompleteness---a combination of package versions that
  satisfy dependency constraints does exist, but the package manager is unable
  to find it.

  In this paper we look back at proposals from dependency solving research
  dating back a few years. Specifically, we review the idea of treating
  dependency solving as a separate concern in package manager implementations,
  relying on generic dependency solvers based on tried and tested techniques
  such as SAT solving, PBO, MILP, etc.

  By conducting a census of dependency solving capabilities in state-of-the-art
  package managers we conclude that some proposals are starting to take off
  (e.g., SAT-based dependency solving) while---with few exceptions---others
  have not (e.g., outsourcing dependency solving to reusable components). We
  reflect on why that has been the case and look at novel challenges for
  dependency solving that have emerged since.
\end{abstract}

\begin{IEEEkeywords}
  software components, dependency solving, SAT solving, package manager,
  separation of concerns
\end{IEEEkeywords}

\section{Introduction}
\label{sec:intro}

Initially introduced in the early 90s, package managers have been used to
support the life-cycle of software components---listing available packages,
installing, removing, and/or upgrading them---for several decades now.
Initially prevalent in UNIX-like software distributions, they have reached peak
popularity during the past decade expanding first to development stacks for
library management---at the time of writing libraries.io~\cite{librariesio2018}
lists more than 30 package managers, most of which are programming
language-specific---and then to final users in various ``app stores'' forms.

One of the key responsibilities of package
managers~\cite{hotswup-package-upgrade} is \emph{dependency solving}. In a
nutshell, a dependency solver takes as input: (1) the current \emph{status} of
packages installed on a given system, (2) a \emph{universe} of all available
packages, (3) a \emph{user request} (e.g., ``install the \texttt{aiohttp}
library''), and (4) explicit or implicit \emph{user preferences} (e.g., ``only
install strictly required packages'' v.~``install all recommended packages
too''). As its output, a dependency solver produces an \emph{upgrade plan},
which is a partially ordered list of low-level actions that should be executed
to reach a \emph{new} status that satisfies the user request; example of such
actions are ``download version 18.2.0 of the \texttt{attr} library'',
``uninstall version 3.5.4 of aiohttp'', and ``install version 3.6.2 of aiohttp
from downloaded zip file''.

Dependency solving is a hard problem in all non-trivial component models. It
has first been shown to be NP-complete in 2006 for expressive dependencies such
as Debian's~\cite{mancinelli2006sat}---which allows version predicates (e.g.,
\texttt{python3-aiohttp >= 3.0.1}), AND/OR logical connectors, virtual
packages, and explicit inter-package conflicts. Intuitively, the difficulty of
dependency solving comes from the fact that it is not enough to explore the
dependency tree of the package you want to install, because you might need
arbitrarily deep backtracking to check \emph{if} a valid solution to the user
request does exist. In formal terms, (Debian's) dependency solving can be
encoded as a SAT solving problem and vice-versa~\cite{garey2002intractability,
  mancinelli2006sat, LeBerreP08}.

More recently~\cite{jss2012-concern} it has been shown that even much simpler
component models induce NP-completeness, it is enough for a package manager to
support multiple package versions and to forbid co-installation of different
versions of the same package (which is almost invariably the case).

The complexity of dependency solving is further increased by the fact that
users generally do not want \emph{a} solution; but rather an \emph{optimal} one
w.r.t.~some criteria, even when they are not stated explicitly. For instance,
when requesting to install \texttt{wesnoth} users generally expect to install
the \emph{minimum amount of additional packages} that allow them to play that
game (also known as the ``minimum install
problem''~\cite{tucker2007opium}). This translate to an optimization problem,
which poses additional challenges on dependency solving implementation.

During the 2005--2015 decade it had been observed how most state-of-the-art
package managers were incomplete (i.e., incapable of proposing a valid upgrade
plan when one existed) and not expressive enough (i.e., not allowing users to
express user preferences to drive the optimization part of dependency solving).
A substantial body of research has been devoted to study dependency solving to
improve the capabilities of package managers, in particular in the framework of
the Mancoosi European research project~\cite{Mancoosi}.

In this paper we look back at one particular proposal~\cite{jss2012-concern}
from back then, that of \emph{treating dependency solving as a separate concern
  in package manager} design and implementation, delegating it to a
specialized, highly-capable dependency solver based on state-of-the-art
constraint solving and optimization techniques.

\begin{figure*}[t]  \centering
  \includegraphics[width=0.75\textwidth]{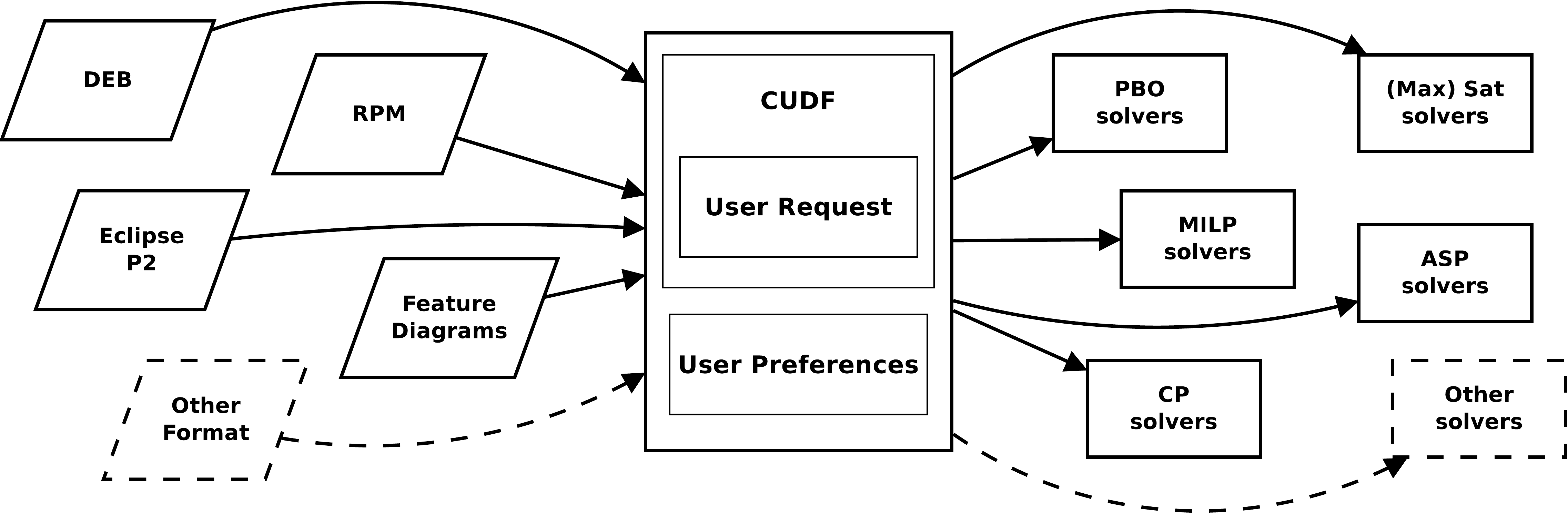}
  \caption{CUDF: a common language to encode dependency solving scenarios
    (figure from~\cite{jss2012-concern})}
  \label{fig:cudf}
\end{figure*}

\paragraph*{Paper structure}
We review the ``separate concern'' proposal in Section~\ref{sec:proposal}; we
conduct a census of dependency solving capabilities for state-of-the-art
package managers (Section~\ref{sec:census}); based on census results we reflect
on what has actually came true of that proposal (Section~\ref{sec:back}); we
conclude considering novel challenges for dependency solving
(Section~\ref{sec:forward}).

\section{Dependency Solving as a Separate Concern}
\label{sec:proposal}

\begin{figure*}[t]
  \centering
  \includegraphics[width=0.8\textwidth]{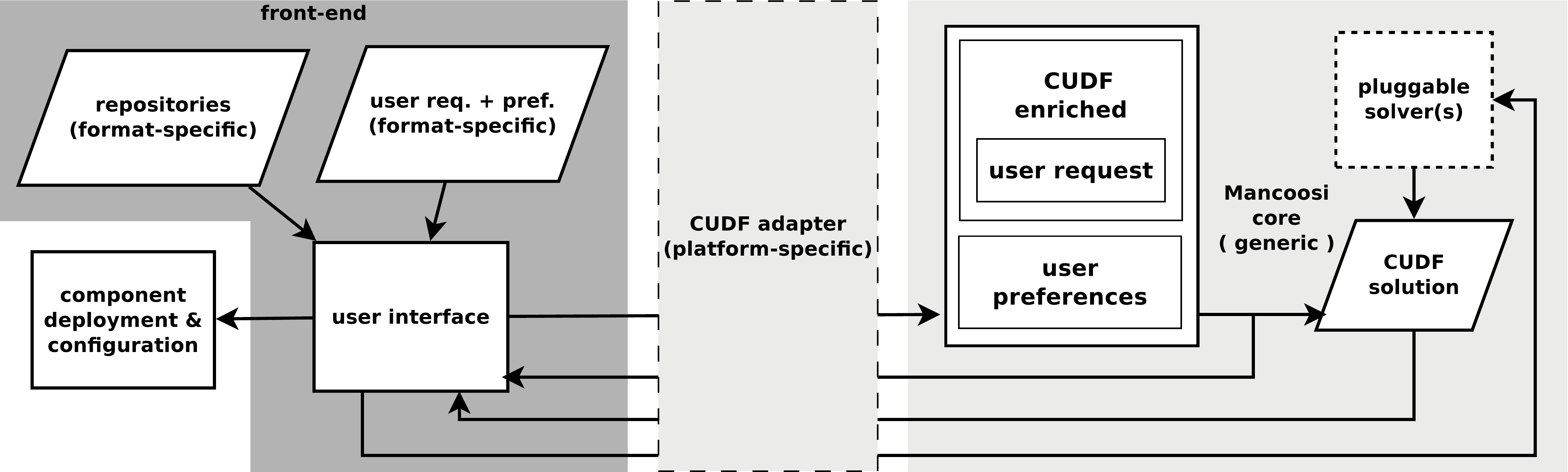}
  \caption{A modular package manager architecture (figure
    from~\cite{infsof2012-mpm})}
  \label{fig:mpm}
\end{figure*}

We can breakdown the research proposal~\cite{jss2012-concern} we are reviewing
into two main claims. The first was that dependency solving should be
expressive. \emph{Expressive} in the sense that dependency expressions should
be powerful (package name and version predicates, conflicts, boolean
connectors, etc.) and that users should have the possibility of expressing
their own optimization criteria to complement built-in ones. To reap the
benefits of such expressivity dependency solvers should be \emph{complete}. And
to that end dependency solver implementations should not be improvised using
ad-hoc heuristics, but rather delegated to specialized solvers based on tried
and tested techniques in constraint solving and optimization.

The second claim was that there is no need to reinvent the dependency solving
wheels over and over again, once for each package manager. We can instead build
capable dependency solvers once (multiple times only if justified by the use of
different techniques or to innovate in neighbor areas), and plug them into
package managers as needed.

To support these claims a formal representation language called CUDF (for
Common Upgradeability Description Format~\cite{cudf2-spec}) was designed, with
the idea of using it as a \emph{lingua franca} between package managers and
solvers, as depicted in \figref{fig:cudf}. According to this view a package
manager facing a dependency solving user request will first translate it to an
\emph{upgrade problem} expressed in CUDF, then invoke a CUDF-enabled dependency
solver on it, which will return a CUDF-encoded solution to the original package
manager. As shown in the \emph{modular package manager} architecture of
\figref{fig:mpm}, only the back and forth CUDF translations are
platform-specific; dependency solvers themselves are package manager agnostic
and hence reusable.

\begin{table}
  \caption{General purpose, CUDF-enable dependency solvers (MISC 2010--2011
    sample participants).}
  \label{tab:misc-solvers}
  \centering
  \begin{tabular}{ll}
    \multicolumn{1}{c}{\textbf{CUDF solver}}
    & \multicolumn{1}{c}{\textbf{technique / solver}}
    \\\hline
    \emph{apt-pbo}~\cite{trezentos10ase} & Pseudo Boolean Optimization \\
    \emph{aspcud}~\cite{aspcud} & Answer Set Programming \\
    \emph{inesc}~\cite{argelich10lococo} & Max-SAT \\
    \emph{p2cudf}~\cite{argelich10lococo} & Pseudo Boolean Optimization / Sat4j~\cite{sat4j} \\
    \emph{ucl} & Graph constraints \\
    \emph{unsa}~\cite{michel10lococo} & Mixed Integer Linear Programming / CPLEX~\cite{cplex} \\
    \hline
  \end{tabular}
\end{table}

As practical evidence of the feasibility of that approach an international
dependency solving competition, called MISC~\cite{jss2012-concern}, has been
run for 3 yearly editions from 2010 to 2012, using CUDF as the input/output
format for participating solvers. The competition has been run on real
dependency solving problems gathered by package manager users (via a submission
system) as well as on randomly generated ones, starting from real-world package
repositories. All data used as input for the competition has been made publicly
available~\cite{MiscData}. As a byproduct of MISC, several CUDF-speaking
general purpose dependency solvers have been released; some examples are shown
in Table~\ref{tab:misc-solvers}.

\section{A Dependency Solving Census}
\label{sec:census}

\begin{table*}[t]
  \centering
  \caption{Dependency solving feature matrix for state-of-the-art package
    managers.}
  \label{tab:census}
\includegraphics[width=\textwidth]{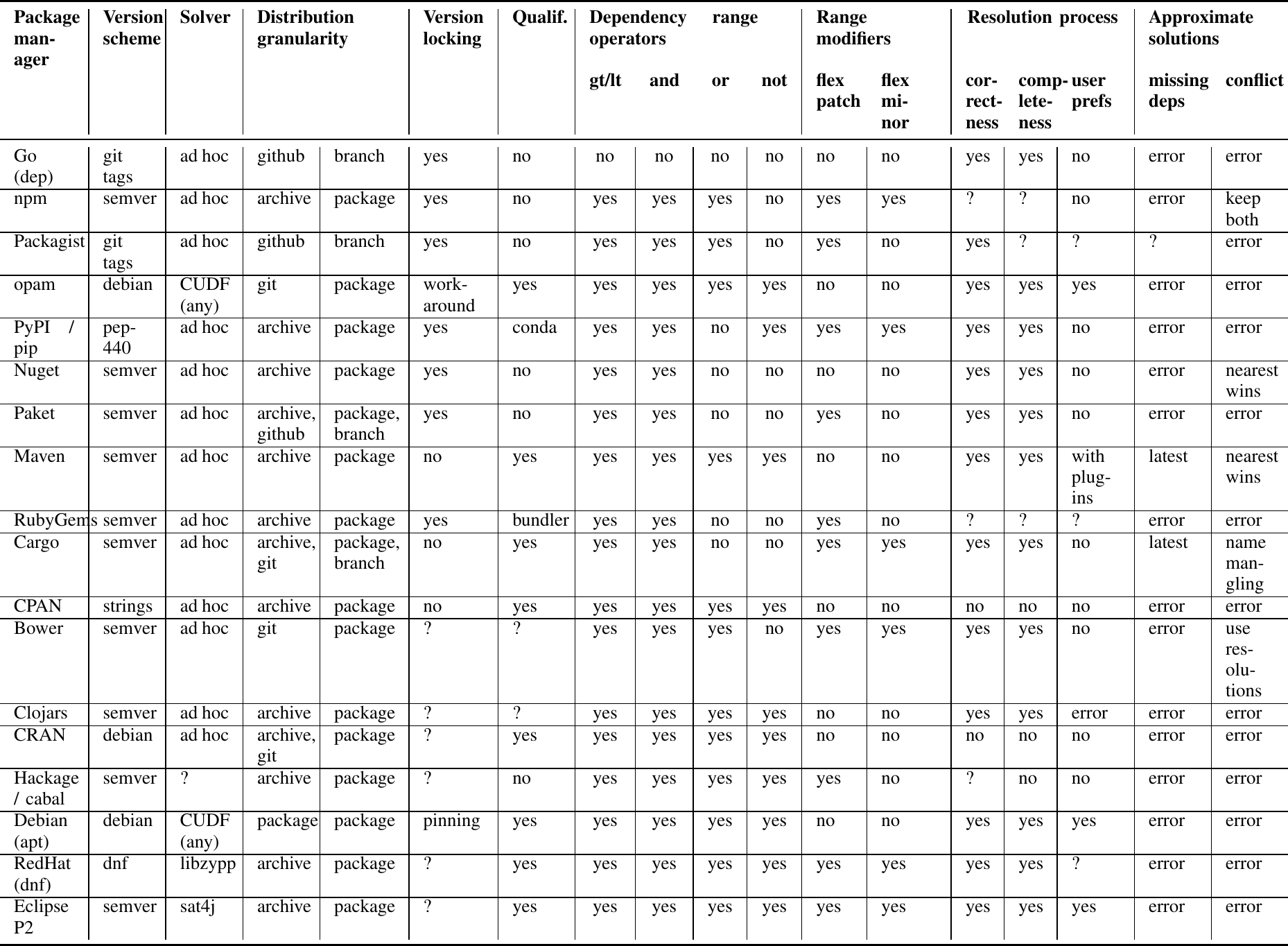}
\end{table*}

Almost a decade later, has this view of expressive, complete, and mutualized
dependency solving become true?

To verify that we have conducted a census of the dependency solving
capabilities of current package managers. We have included in the census major
language-specific package managers from libraries.io~\cite{librariesio2018} as
well as package managers from notable Free/Open Source Software (FOSS)
distributions and platforms, such as Debian, RedHat and Eclipse.

Census results are summarized in Table~\ref{tab:census}. For each package
manager we considered the following dimensions:

\paragraph*{Versioning scheme} How does the package manager specify versions
for the artifacts it manages? Common versioning schemes include semantic
versioning (\textsf{semver}) and its derivatives, where a version is identified
by a quadruplet \texttt{major.minor.patch.qualifier}, where each qualifier
specifies an order. Other schemes include Debian's version spec
(\textsf{debian}) and using free form strings with no ordering semantics
(\textsf{git tags}, \textsf{strings}).

\paragraph*{Distribution} How are packages distributed? Most package managers
use centralized \textsf{archive}s, whereas a new trend is to use
\textsf{github} as a distribution platform in addition to collaboration.

\paragraph*{Granularity} What is the minimal unit that can be versioned?  Most
dependency managers version artifacts at the \textsf{package} level, but some,
notably those that support package distribution over \textsf{github} also allow
versioning of repository branches.

\paragraph*{Version Locking} Does the package manager support locking the
results of a package resolution? Most package managers enable this option, to
help developers maintain reproducible builds.

\paragraph*{Qualifiers} Does the package manager support selecting specific
dependencies based on external build configurations? One such typical example
is the inclusion of test runner dependencies only when running tests. Many
package managers enable this feature to minimize the set of dependencies in
specific environments.

\paragraph*{Dependency range operators} What levels of expressivity does the
package manager range specification language enable? Package managers that use
semantic versioning (or other types of hierarchical versioning) enable users to
specify ranges of dependency versions a package depends upon. For example, a
package might depend on all patch versions of an artifact version \texttt{4.3};
this can be expressed as a range: \texttt{>=~4.3.*}. To express more complex
scenarios, many package managers allow boolean operators on ranges.

\paragraph*{Range modifiers} Even more complex scenarios might arise with
dependency ranges: what if a developer wants to express a constraint such as
``update to all new minor versions, but not to the next major one''. Range
modifiers enable developers to anticipate new patch (\textsf{flex patch}) or
new minor (\textsf{flex minor}) versions without having to explicitly modify
their project's manifest files.

\paragraph*{Resolution process} We consider the following facets of package
managers approaches to dependency solving:
\begin{itemize}

\item \emph{Correctness:} Will the package manager always propose solutions
  that respect dependency constraints?

\item \emph{Completeness:} Will the package manager always find a solution if
  one exists?

\item \emph{User preferences:} Can the user provide custom optimization
  criteria to discriminate among valid solutions? For example, in order to
  minimize/maximize the number of packages matching stated
  characteristic~\cite{mooml-iwoce-2009} or to veto certain packages.

\end{itemize}

\paragraph*{Approximate solutions} When a solution cannot be found, some
package manager may try to proceed anyway by relaxing some constraints.
\begin{itemize}

\item \emph{Missing dependencies:} When a dependency version constraint cannot
  be satisfied, most package managers will report an error, while some (e.g.,
  Cargo and Maven) will ignore the error and install the latest available
  version.

\item \emph{Conflicts:} When the transitive closure of a dependency resolution
  includes more than one version of the same artifact, most package managers
  will bail out with an error, as no valid solution exists. Some package
  managers on the other hand will force the installation to complete
  nonetheless: Cargo rewrites the conflicting symbol names to enable multiple
  versions of libraries to co-exist; others select the version that is closer
  to the root of the dependency tree of the package whose dependencies are
  being resolved.

\end{itemize}

Among the various features listed above, user defined preferences for driving
dependency resolution appear to be the least known, hence we provide here a few
examples to illustrate what they look like and how they are used.

The \texttt{opam} package manager for the OCaml programming language offers the
user a rich set of preferences,\footnote{See
  \url{https://opam.ocaml.org/doc/External_solvers.html} for full details.}
here is an example:
{\footnotesize
\begin{verbatim}
 opam install merlin --criteria="-changed,-removed"
\end{verbatim}
}
which requests to install \texttt{merlin}. Since this is a development tool,
the user does not want its installation to impact other libraries installed in
the system that might be also used as build dependencies of the project. To
this end, the \texttt{-changed,-removed} preferences indicate that, among all
possible solutions, we prefer the one that minimizes changes to the system, and
minimizes removal of other packages.

\section{Discussion}
\label{sec:back}

The first observation about census findings (Table~\ref{tab:census}) is that,
almost 15 years after the seminal work dependency solving NP-completeness, a
significant set of package managers rely on robust, specialized solvers, able
to support correct and complete dependency solving---e.g., Eclipse uses P2,
built on top Sat4J~\cite{sat4j}, SUSE and RedHat use \texttt{libsolv} (itself
based on the
libzypp\footnote{\url{https://en.opensuse.org/openSUSE:Libzypp_satsolver}} SAT
solver), while Debian and Opam can use any external CUDF solver. This is good
news: the importance of using complete dependency solvers seems now well
acknowledged and it seems to be common knowledge that this entails leveraging
solver technologies like SAT, MaxSAT, PBO, ASP or MILP, instead of ad-hoc
dependency graph traversals. We consider that a significant part of the first
claim of~\cite{jss2012-concern} actually made it through.

On the other side, it seems that only Opam has embraced~\cite{cudf-ocaml-2014}
the ``separation of concern'' approach advocated in~\cite{jss2012-concern},
with \texttt{apt-get} somewhat halfway through, as it offers access to external
solvers only as an option. There are several factors that may explain this
limited success: some are technical, others are of social nature.

From the technical point of view, we notice two issues. First, the CUDF format
has some shortcomings. While it is very well adapted for package managers that
use versioning and dependency schemes similar to the Debian ones, it does not
support natively dependency constraints involving qualifiers (used by Eclipse
P2) or non overlapping version intervals (npm)---they \emph{can} be supported,
but at the cost of additional complexity in the CUDF adapter.  Second, while
relying on one or more external solvers may be a smart choice in the long
run,\footnote{This is shown by the recent switch made in Opam from the
  \texttt{aspcud} solver to \texttt{mccs}, triggered by performance issues that
  only showed up with the growing number of existing packages.} it introduces
an external dependency in a key component, the package manager, that needs to
be properly catered for.  These two aspects have likely reduced the buy-in on
relying on third party CUDF solvers.

As for the social aspects, a broad adoption of the ``separation of concern''
approach would mean convincing \emph{not one community, but many}, to adapt the
architecture of one of their key tools and accept to rely a \emph{common
  standard} on which they would have individually little leverage. This is a
significant social challenge, and it is understandable that many preferred to
retain full control on their package manager, and just hard-wire in it a
specific solver, especially when one written in the same programming language
was available.

Hence we believe that it is already a significant success to see the proposed
approach embraced in full by the Opam package manager, which is also the only
one offering full support for flexible user preferences. The direct implication
in the Opam/OCaml community of some of the proponents of~\cite{jss2012-concern}
has surely been an important adoption factor too. ``If you build it, they will
come'' is not always enough; broad adoption also needs to actually go out of
your way (and role) to \emph{make} the needed adaptations and provide concrete
evidence of the conveyed advantages.

\vspace{-0.4em}
\section{Outlook}
\label{sec:forward}

``Dependency hell'' is a colloquial term denoting the frustration resulting
from the inability to install software due to complicated dependencies. From
the review we conducted one cannot conclude that the problem is solved.
However, the situation significantly improved w.r.t.~less than a decade
ago. Several package managers are both correct and complete---the two
properties that contribute the most to addressing the dependency hell---and the
reinvention of dependency solving wheels has been avoided in at least a few
notable cases. All in all, it seems that good dependency solving practices are
spreading, which makes us hopeful for a better future.

Novel depdency management approaches have emerged since the proposals reviewed
in this paper. On the one hand, containerization and virtual environments have
gained significant traction; functional package managers~\cite{nixos, guix}
have become more popular, due to analogies with container technology and a
surge in the interest for scientific and build reproducibility. These
approaches share the ability to create separate package namespaces on-the-fly,
allowing to deploy side-by-side packages that would be incompatible in a shared
namespace. This has alleviated the need for correct and complete dependency
solving, but we speculate it will not for long---the recent
announcement\footnote{\url{https://github.com/python/request-for/blob/master/2020-pip/RFP.md}}
that PyPI/pip, a software ecosystem in which virtual environments are really
popular, is finally going to implement proper dependency solving seems to be a
step in the right direction.

Novel challenges are emerging on the front of dependency \emph{auditing}. For
example, there is no way for developers to know whether a security issue
affecting a dependency is also affecting their programs. Licensing
incompatibilities cannot be easily detected either, even though most packages
come with accompanying license metadata.
The root cause behind those issues is that the finest granularity in package
management is still the package, whereas software reuse happens at finer levels
(e.g., modules, functions, etc.)~\cite{Mycroft2017}. This discrepancy leads to
lost opportunities. The construction of inter-package call graphs, as envisaged
by the FASTEN~\cite{fasten} project, may unlock several new package manager
features, such as precise tracking of security and licensing incompatibility
issues, data-driven API evolution, and several others.

\section*{Acknowledgements}

This work has been partially funded by the FASTEN project, part of the European
Commission H2020 program (contract: 825328).

\end{document}